\documentclass[cits]{PoS}
\usepackage{graphicx}

    \newcommand{\ev}{\hbox{ eV}}

    \newcommand{\gev}{\hbox{ GeV}}
    \newcommand{\tev}{\hbox{ TeV}}
    \newcommand{\cm}{\hbox{ cm}}
\newcommand{\onetev}{1-TeV scale}

\newcommand{\ewgg}{\ensuremath{\mathrm{SU(2)_L }\otimes \mathrm{U(1)}_Y}}
\newcommand{\ie}{{\em i.e.}}
\newcommand{\abs}[1]{\left| #1\right|}
\newcommand{\PoS}[1]{in proceedings of
\emph{10th International Workshop on Neutrino Factories, Super beams and Beta beams (NuFact08)},
June 30 -- July 5 2008, Valencia, Spain \pos{PoS(NUFACT08)#1}}

\title{Closing talk}

\ShortTitle{Closing talk}

\author{\speaker{CHRIS QUIGG}
\\  Institut f\"{u}r Theoretische Teilchenphysik, 
Universit\"{a}t Karlsruhe, 
D-76128 Karlsruhe, Germany\\
       E-mail: \email{chris.quigg@gmail.com}}

\abstract{Nufact08 is the tenth in a series of workshops started in 1999, whose main goal is to understand options for future neutrino-oscillation experiments to attack the problems of the neutrino mass hierarchy and \textsf{CP} violation in the leptonic sector. I present a very brief review of what we know and what we would like to know about neutrino mass, mixing, and flavor change. I consider the interplay between neutrino physics and forthcoming information from the Large Hadron Collider. I comment on a decade's progress and offer some context for work that lies ahead.}

\FullConference{10th International Workshop on Neutrino Factories, Super beams and Beta beams\\
                June 30 - July 5 2008\\
                Valencia, Spain}

\begin{document}

\section{Introductory Remarks}
We have enjoyed a lively and productive week, with informality and engagement promoted by the workshop's setting in Valencia's Botanical Garden and an abundance of information and ideas brought by many committed machine builders, experimentalists, and theorists~\cite{Pilar,ChrisW,Hayato,WG3,Flavio}. Enormously impressive progress has been reported this week. I have been repeatedly struck by the mounting coherence of the experimental results, the breadth of the scientific opportunities, and the creativity of machine designers. We have found much incentive to stretch our minds and to think ``blue-sky'' thoughts about machines, detectors~\cite{Soler}, baselines, and strategies. On occasion, we have even found the discipline to look closely at the physics goals and ask how little we might require to extract the information we need. We return to our institutions with a lot to think about, and with a good list of homework assignments for the coming year. 
In addition to the material in \textit{These Proceedings,} we and our colleagues will be able to mine the slides presented during the workshop, which can be found online at \href{http://ific.uv.es/nufact08/}{\tt ific.uv.es/nufact08/}.

In my opening talk at the first Nufact in Lyon~\cite{Quigg:1999ap}, I offered a framework to guide our work: {\it
As we begin this workshop, it seems to me that we should keep in mind 
four essential questions:
	Is a Neutrino Factory feasible? 	At what cost? 
	How soon? 	What R\&D must we do to learn whether we can make the neutrino 
	factory a reality?
The answers to these questions will be influenced by what we want the 
neutrino factory to be.  To decide that, we need to consider another 
set of questions:
	What do we want to know about neutrino masses and mixings now \ldots 
	in five years \ldots in ten years? 	Is a neutrino factory the best way---or the only way---to provide 
	this information?  	What (range of) beam parameters should the neutrino factory 
	offer?  What detectors are needed to carry out the physics program of a 
neutrino factory?  It seems that distant detectors must weigh several 
kilotonnes and ideally should identify electrons, muons, and taus---and 
measure their charges.  Are all these characteristics essential?  How 
should the prospect of a neutrino factory influence the detectors we 
build now?}

We have learned a great deal over a decade from the operation of modern neutrino beams~\cite{Edda}, and are on the cusp of learning still more from 
the MICE experiment~\cite{mice} and other technology demonstrations~\cite{Bross,Kirk}. The search for new tools today includes beta beams~\cite{Elena}, super beams~\cite{ProjectX}, and reactors~\cite{Lhuillier}, along with the dream of a muon collider~\cite{BobP}, which spawned the modern notion of a neutrino factory.
It is essential now, as it was then,  to consider the scientific issues with an eye to both the intrinsic interest in neutrino properties and interactions and also 
the evolving place of neutrino studies within contemporary particle 
physics.  I set the context in 1999 in terms of ten Questions of Identity: {\it
Do neutrinos oscillate?
What are the neutrino masses?
Is neutrino mass a sign of (nontrivial) physics beyond the standard model?
Does the evidence require more than three neutrino species?
Can we find evidence for (or against) a sterile neutrino?
Could neutrino masses be special?
How could light sterile neutrinos arise?
Are neutrino mixing angles large? maximal?
Do neutrino masses probe large extra dimensions?
Can we detect \textsf{CP} violation in neutrino mixing? } We have answered a few of these questions, enriched our understanding of others, and been led to still others. I give a revised list at the end of this talk.

\begin{figure}[tb]
\begin{center}
\centerline{\includegraphics[width=0.4\textwidth]{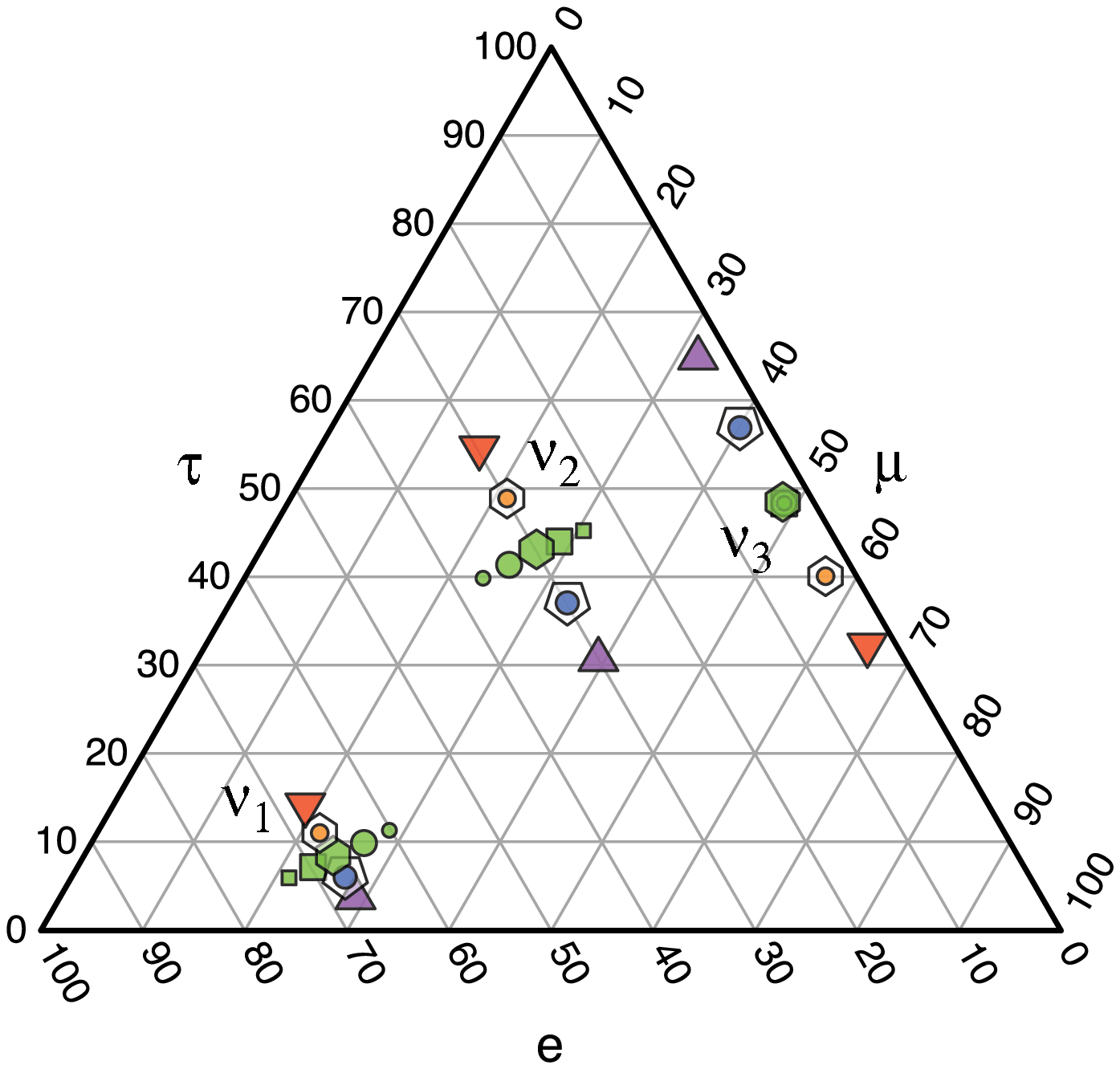} \quad\includegraphics[width=0.4\textwidth]{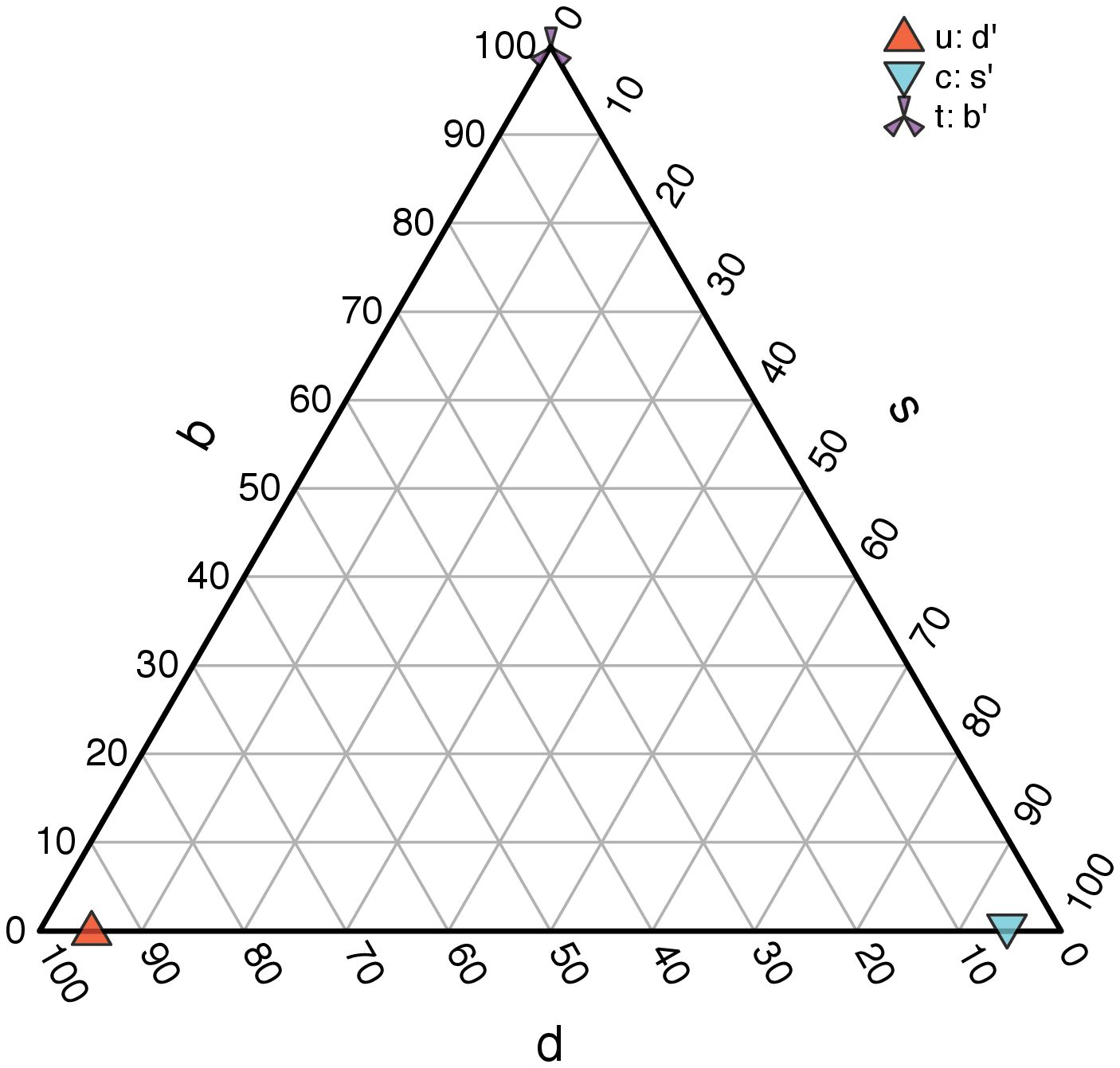}}
\caption{Left pane: $\nu_e, \nu_\mu, \nu_\tau$ flavor content of the neutrino mass eigenstates $\nu_1, \nu_2, \nu_3$. The green hexagons denote central values, with $\delta = 0$ and $\theta_{13} = 10^\circ$. Variations in the atmospheric angle $\theta_{23}$ are indicated by the points arrayed roughly parallel to the $\mu$ scale. Variations in the solar angle $\theta_{12}$ are depicted by the green symbols arrayed roughly perpendicular to the $\mu$ scale. Right pane: $d, s, b$ composition of the quark flavor eigenstates $d^\prime$ (red $\bigtriangleup$), $s^\prime$ (green $\bigtriangledown$), $b^\prime$ (violet tripod).
\label{fig:flavormixing}}
\end{center}
\end{figure}

Neutrinos do change flavor, and the established phenomena can be interpreted in terms of the mixing of three flavors of neutrinos composed of three mass eigenstates with masses $m_1$, $m_2$, $m_3$.\footnote{See Andr\'{e} de Gouv\^{e}a's talk~\cite{Gouv} for reasons to consider more than three species.}
It is conventional to factor the neutrino mixing matrix as~\cite{Boris} 
\begin{equation}
\mathsf{
U = \left(
\begin{array}{ccc}
1 & 0 & 0 \\
0 & c_{23} & s_{23} \\
0 & -s_{23} &  c_{23}
\end{array}
\right)
\left(
\begin{array}{ccc}
c_{13} & 0 & s_{13}e^{-i\delta} \\
0 & 1 & 0 \\
-s_{13}e^{i\delta} & 0 & c_{13}
\end{array}
\right)
\left(
\begin{array}{ccc}
c_{12} & s_{12} & 0 \\
-s_{12} & c_{12} & 0 \\
0 & 0 & 1
\end{array}
\right)
}\cdot \left( \begin{array}{ccc}
1 & 0 & 0 \\
0 & e^{i\varphi_2} & 0 \\
0 & 0 & e^{i\varphi_3}  
\end{array}
\right)\;,
\end{equation}
where we abbreviate $s_{ij} = \sin\theta_{ij}$, $c_{ij} = \cos\theta_{ij}$, and $\delta$ is a \textsf{CP}-violating phase. The final factor, containing two Majorana phases, is present only if the neutrino is its own antiparticle. Our current knowledge~\cite{Hitoshi,Concha,Niki,Kajita} of the three mixing angles restricts the ``solar angle'' $30^\circ \lesssim  \theta_{12} \lesssim  38^\circ$, the ``atmospheric angle'' $35^\circ \lesssim  \theta_{23} \lesssim  55^\circ$, and the ``small angle'' $ \theta_{13} \lesssim  10^\circ$. The $\textsf{CP}$-violation phase $\delta$ is unconstrained.
These parameter ranges lead to the flavor content of the neutrino mass eigenstates depicted in the left pane of Figure~\ref{fig:flavormixing}, 
where central values (fixing $\delta = 0$ and $\theta_{13} = 10^\circ$) are indicated by the green hexagons. We observe that  $\nu_3$ consists of nearly equal parts of $\nu_{\mu}$ and $\nu_{\tau}$, perhaps with a trace of $\nu_e$, while $\nu_2$ contains similar amounts of $\nu_e$, $\nu_{\mu}$, and $\nu_{\tau}$, and $\nu_1$ is rich in $\nu_e$, with approximately equal minority parts of $\nu_{\mu}$ and $\nu_{\tau}$. The observed structure of the neutrino mixing matrix differs greatly from the pattern of the more familiar (Cabibbo--Kobayashi--Maskawa) quark mixing matrix, which is displayed graphically in the right pane of Figure~\ref{fig:flavormixing}.

\begin{figure}[bt]
\begin{center}
\centerline{\includegraphics[width=0.4\textwidth]{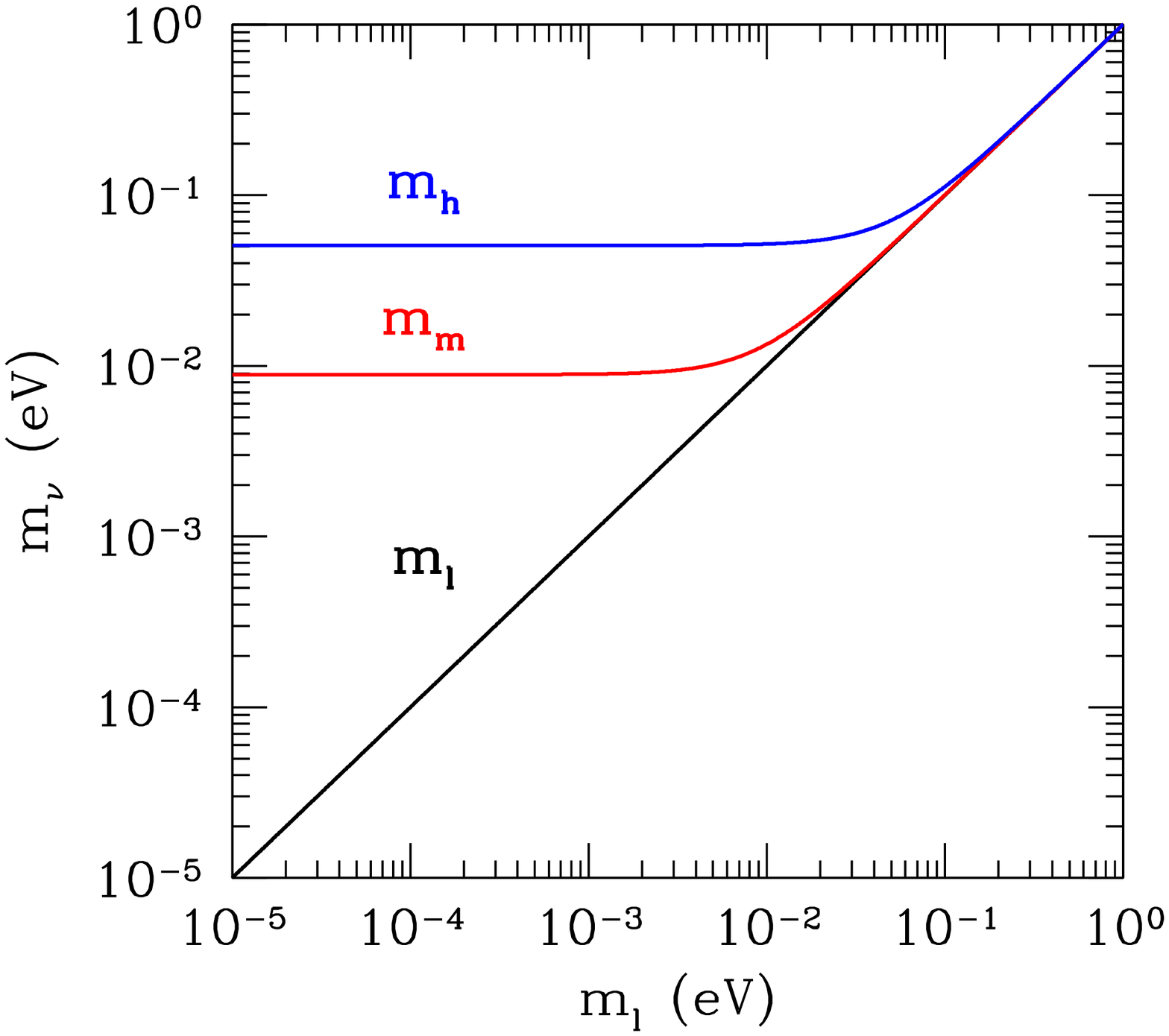} \quad \includegraphics[width=0.4\textwidth]{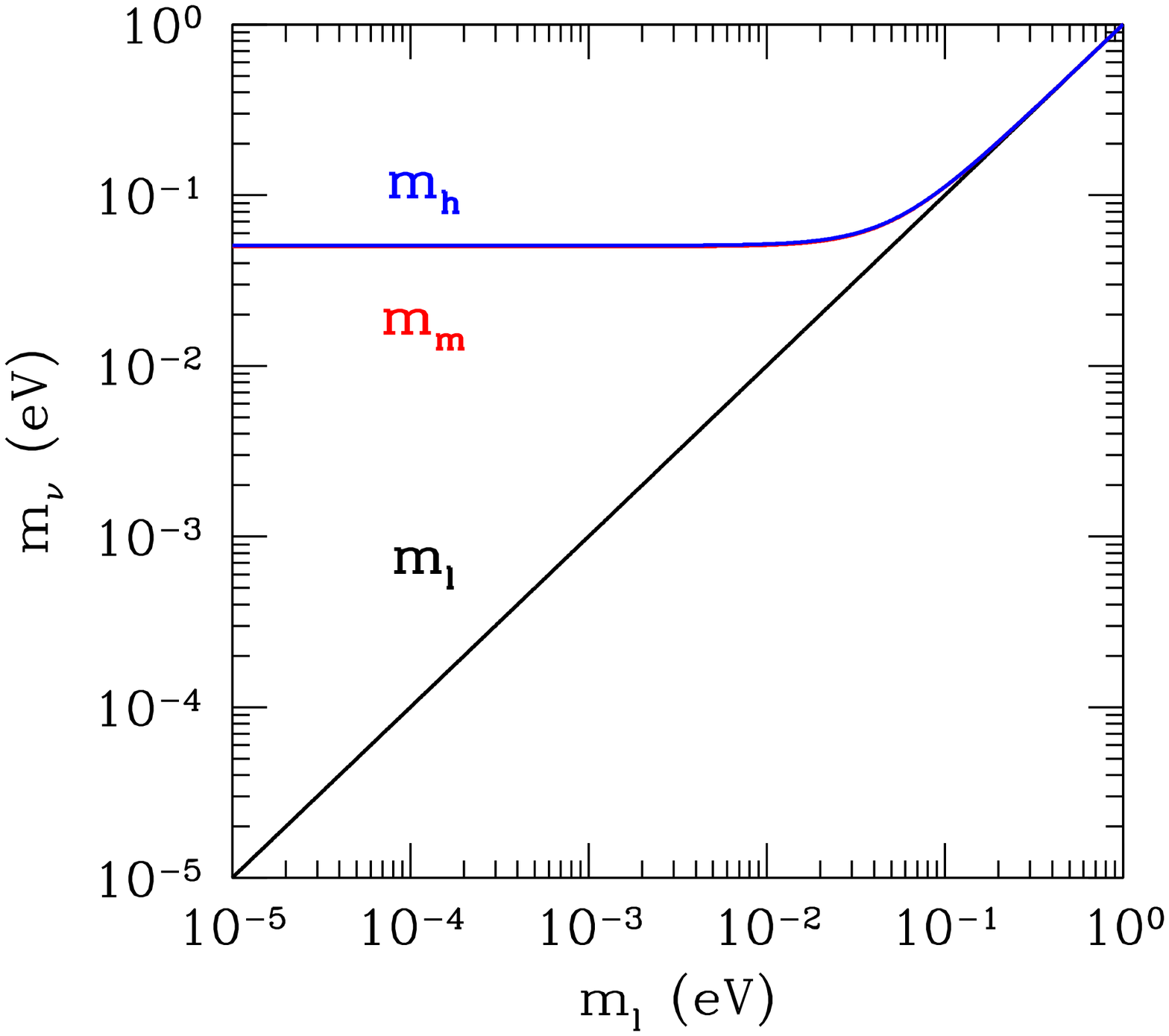}}
\caption{Favored values for the light, medium, and heavy neutrino masses $m_{\ell}$, $m_{\mathrm{m}}$, $m_{\mathrm{h}}$, as functions of the lightest neutrino mass  in the three-neutrino oscillation scenario for the normal (left pane) and inverted hierarchy (right pane). We take the solar mass-squared difference to be $\Delta m^2_{\odot} = m_2^2 - m_1^2 = 7.9\times 10^{-5}\ev^2$, and the atmospheric
$ \Delta m^2_{\mathrm{atm}} = |m_3^2 - m_1^2| = 2.5\times 10^{-3}\ev^2$.
\label{fig:neutrinomasses}}
\end{center}
\end{figure}

\begin{figure}[tb]
\begin{center}
\includegraphics[width=0.54\textwidth]{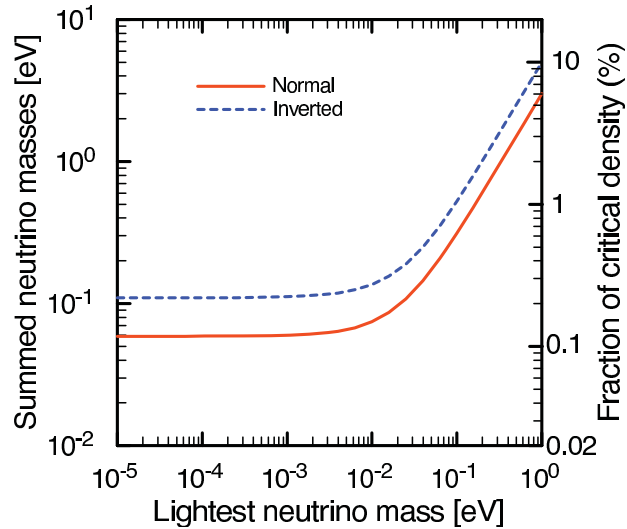}
\caption{Contributions of relic neutrinos to the mass density of the Universe, as functions of the mass of the lightest neutrino, for the normal (solid line) and inverted (dashed line) mass hierarchies.
\label{fig:neuDM}}
\end{center}
\end{figure}

 The atmospheric and solar neutrino experiments, with their reactor and accelerator complements, have partially characterized the neutrino spectrum in terms of a closely spaced solar pair $\nu_1$ and $\nu_2$, where $\nu_1$ is taken by convention to be the lighter member of the pair, and a third neutrino, more widely separated in mass. We do not yet know whether $\nu_3$ lies above (``normal hierarchy'') or below (``inverted hierarchy'') the solar pair, and experiment has not yet set the absolute scale of neutrino masses. Figure~\ref{fig:neutrinomasses} shows the normal and inverted spectra as functions of assumed values for the mass of the lightest neutrino.
With our (partial) knowledge of neutrino masses, we can estimate the contribution of neutrinos to the density of the current universe. The left-hand scale of Figure~\ref{fig:neuDM} shows the summed 
neutrino masses $m_1 + m_2 + m_3$ for the normal and inverted hierarchies, as functions of the lightest neutrino mass. The neutrino oscillation data imply that $\sum_i m_{\nu_i} \gtrsim 0.06\ev$ in the case of the normal hierarchy, and $\sum_i m_{\nu_i} \gtrsim 0.11\ev$ in the case of the inverted hierarchy.
Indirect inferences from cosmological observations promise to reach a sensitivity of $\sum_i m_{\nu_i}\approx 0.03\ev$~\cite{Sergio,Silvia}, so we have the prospect of a concordance or a contradiction that would suggest the need to revise our thinking about the evolution of the universe. The KATRIN tritium beta-decay experiment~\cite{katrin} aims for a sensitivity of $0.2\ev$ for the kinematically determined mass of the neutrino emitted in beta decay.

Using the calculated number density of $56\cm^{-3}$ for each $\nu$ and $\bar{\nu}$ flavor in the current universe, we can deduce the neutrino contribution to the mass density, expressed in units of the critical density, as $\rho_{\mathrm{c}} \equiv 3H_0^2/8\pi G_{\mathrm{N}} = 1.05 h^2 \times 10^4\ev\cm^{-3} = 5.6 \times 10^3\ev\cm^{-3}$, where $H_0$ is the Hubble parameter now, $G_{\mathrm{N}}$ is Newton's constant, and
I have taken the reduced Hubble constant to be $h = 0.73$. This is measured by the right-hand scale in Figure~\ref{fig:neuDM}. We find that neutrinos contribute $\Omega_\nu \gtrsim (1.2, 2.2) \times 10^{-3}$ for the (normal, inverted) spectrum, and no more than 10\% of critical density, should the lightest neutrino mass approach $1\ev$. The condition for neutrinos not to overclose the Universe is
$ \sum_i m_{\nu_i} \lesssim 50\ev$,
so long as neutrinos are stable on cosmological time scales and the expected neutrino density is not erased by interactions beyond the standard electroweak theory.\footnote{See~\cite{Quigg:2008ab} for background and extensive references to the cosmic-neutrino literature.}

\section{Neutrino Physics in the LHC Era}
The study of neutrino properties is often seen as standing apart from collider physics, because of the suspicion that the origin of neutrino mass implicates energy scales much higher than can be studied directly.\footnote{For a survey of the history and present status of the see-saw mechanism, and a look at alternatives, see~\cite{Seesaw}.} But there is good reason to believe that experiments at the LHC will have important consequences for the way we think about neutrino physics~\cite{Pilaf}, including the possibility that neutrino masses are set on the \onetev\ and will be correlated with new phenomena observed there~\cite{Chen:2006hn}. Let us take a moment to recall why the \onetev\ is privileged for our understanding of the electroweak theory and what new insights we expect from LHC experiments~\cite{Quigg:2007dt}.

The electroweak theory does not give a precise prediction for the mass of the Higgs boson, but a partial-wave-unitarity argument~\cite{Lee:1977eg} leads to a conditional upper bound on the Higgs-boson mass that identifies a key target for experiment. 
We compute the 
amplitudes for gauge-boson scattering at high energies, and make
a partial-wave decomposition.
 Four channels are interesting:
$W_L^+W_L^-$, $Z_L^0Z_L^0$, $HH$, and $ HZ_L^0 $,
where the subscript $L$ denotes the longitudinal polarization
states. For these, the $s$-wave amplitudes are all asymptotically
constant (\ie, well-behaved) and  
proportional to $G_{\mathrm{F}}M_H^2$ in the high-energy 
limit. Requiring that the largest eigenvalue respect the 
partial-wave unitarity condition $\abs{a_0}\le 1$ yields
$M_H^2 \le 8\pi\sqrt{2}/3G_{\mathrm{F}} \approx 1\tev^2$
as a condition for perturbative unitarity.

If the bound is respected, weak interactions remain weak at all
energies, and perturbation theory is everywhere reliable. If the
bound is violated, perturbation theory breaks down, and (in the standard-model framework) weak
interactions among $W^\pm$, $Z$, and $H$ become strong on the \onetev.
Features of strong interactions at GeV energies
would then characterize electroweak gauge boson interactions at
TeV energies. More generally, we conclude that new phenomena are to
be found in the electroweak interactions at energies not much larger
than 1~TeV.

To better understand why getting to the root of electroweak symmetry breaking is important for our conception of nature, consider what the world would be
like absent anything resembling the Higgs mechanism. Quarks and leptons would
remain massless  if the electroweak symmetry remained
manifest. Quantum Chromodynamics would operate as usual, confining
 the (massless) color-triplet quarks into color-singlet
hadrons, with very little change in the masses of those stable
structures.  QCD does something more: it hides the electroweak symmetry~\cite{Weinstein:1973gj}! 
In a world with massless up and down quarks, QCD exhibits a global $\mathrm{SU(2)_L\otimes SU(2)_R}$ \textit{chiral symmetry} that treats the left-handed and right-handed quarks
as separate objects. As we approach low energy from above, that chiral
symmetry  is spontaneously broken.  The resulting communication between the
left-handed and right-handed worlds engenders a breaking of the
electroweak symmetry: $\mathrm{SU}(2)_{\mathrm{L}}\otimes \mathrm{U}(1)_{Y}$
becomes $\mathrm{U}(1)_{\mathrm{em}}$, and the gauge bosons are the massless photon and massive $W^\pm$ and $Z^0$. This is not a satisfactory theory of the weak interactions, because the scale of electroweak symmetry breaking is measured by the pion lifetime---the coupling of the axial current to the vacuum. The amount of mass acquired by the $W$ and $Z$ is too small by a factor of 2500. The fermions remain massless, at least in first approximation.

 The familiar spectrum of hadrons persists, but with a crucial difference. The proton now outweighs the neutron, because its greater electrostatic self-energy is not overcome by a mass difference between down and up quarks. Very rapid beta decay, $p \to n e^+ \nu_e$, means that the
lightest nucleus is one neutron; \textit{there is no hydrogen atom.} Light elements created in the early universe might persist until the present, but the Bohr radius of a ``massless'' electron would be infinite. In such a world, there is nothing
we would recognize as an atom, there is no valence bonding,
there are no stable composite structures like the solids and liquids of our everyday experience. Why our world is as we find it will become much clearer once we have understood the mysterious new force that hides the electroweak symmetry.

What form might the answer take?  It could be the Higgs
mechanism of the standard model (or a supersymmetric elaboration),
a force of a new character based on interactions of an elementary scalar. Or it could arise from a new gauge force, perhaps acting on undiscovered constituents. It might be a residual force that arises from the strong dynamics among
the weak gauge bosons. Or perhaps electroweak symmetry breaking is an echo of extra spacetime dimensions. Experiment will show us the path
Nature has taken. An essential first step is to find the Higgs boson and to learn its properties. \textit{Is it there? How many Higgs bosons are there? What are its quantum numbers? Does the Higgs boson generate mass only for the gauge bosons, or does it also give mass to femions? How does the Higgs boson interact with itself?}

The \ewgg\ electroweak theory points to a Higgs-boson mass below $1\tev$ or to other new physics on the \onetev. If there is a light Higgs boson, as suggested by precision electroweak measurements~\cite{ewwg}, the theory does not explain how the 
scale of electroweak symmetry breaking is maintained in the presence 
of quantum corrections.   Beyond the classical approximation, scalar mass parameters receive quadratically divergent
quantum corrections from loops that contain standard-model particles.
In order for the mass shifts 
induced by quantum corrections to remain under control, either nature is exquisitely fine-tuned or some new physics must intervene at an energy not far above the \onetev. Popular among the speculations about new physics include supersymmetry~\cite{Martin:1997ns}, dynamical symmetry breaking~\cite{Hill:2002ap}, and ``little Higgs''~ \cite{Schmaltz:2005ky,Maxim} or ``Higgless''~\cite{He:2007ge}
composite models.

We have yet another indication that new phenomena should be present on the \onetev. An appealing interpretation of the evidence that dark matter makes up roughly one-quarter of the energy density of the universe~\cite{Hinshaw:2008kr} is that dark matter consists of  thermal relics of the big bang, stable---or exceedingly long-lived---neutral particles. If such a particle couples with weak-interaction strength, then generically the observed dark-matter density results if the mass of the dark-matter particle lies between approximately $100\gev$ and $1\tev$~\cite{Bertone:2004pz}. Typically, scenarios to extend the electroweak theory and resolve the hierarchy problem---whether based on extra dimensions, new strong dynamics, or supersymmetry---entail dark-matter candidates on the \onetev. One aspect of the great optimism with which we particle physicists contemplate the explorations under way at Fermilab's Tevatron and soon to be greatly extended at CERN's Large Hadron Collider is a strong suspicion that many of the outstanding problems of particle physics and cosmology may be linked---and linked to the \onetev. Dark matter is a perfect example.

Aside from the hierarchy problem, we believe the standard model to be incomplete because it is characterized by a great number of parameters. Of the twenty-six (or more) apparently arbitrary parameters of the standard model, three ($\alpha_s$, $\alpha_{\mathrm{em}}$, $\sin^2\!\theta_W$) set the strength of gauge couplings, two define the scalar potential, and one sets the vacuum phase of QCD. All the rest are connected with quark and lepton flavor: 6 quark masses, 3 quark mixing angles, and 1 \textsf{CP}-violating phase, plus 3 charged-lepton masses, 3 neutrino masses, 3 leptonic mixing angles, 1 \textsf{CP}-violating phase, and---if the neutrino is its own antiparticle---2 Majorana phases. We cannot know in advance which of these might have fundamental significance and which might be environmental, but correlating new particles and forces observed at the LHC with virtual effects in flavor physics is sure to provide important insights. Flavor physics may well be where we first observe, or successfully diagnose, the break we anticipate in the standard model. I look forward to immensely productive conversations among LHC discoveries, neutrino advances, other high-sensitivity accelerator experiments, and astro/cosmo/particle observations.
Many examples of possible connections were explored in the Nufact08 working groups.
Lifting the electroweak veil around 1 TeV should help us to see the problem of identity (flavor) and the challenges of other scales more clearly.

Completing and extending the electroweak theory is not the only task before us. I believe that a growth industry will be the search for new physics \textit{within} the standard model: phenomena implied by the standard model, but too subtle to have attracted our notice---either theoretical or experimental---until now. A famous example of new physics hiding within the electroweak theory is the nonperturbative violation of baryon number mediated by sphalerons. Of current interest is the (theoretical) identification of a $Z\gamma\omega$ anomaly-mediated neutrino-photon interaction in the presence of baryons~\cite{Harvey:2007ca}, which might help us assess the low-energy excess of electromagnetic energy observed by the MiniBooNE experiment~\cite{AguilarArevalo:2007it}. This kind of deeper look within the standard model provides added motivation for experiments to measure and understand neutrino cross sections (Miner$\nu$a) and hadroproduction (HARP, MIPP, SciBooNE) at low energies~\cite{Cata}.

\section{Outlook}
To close, I would like to update the list of questions with which we began.\footnote{A more comprehensive discussion, with policy recommendations, appears in~\cite{Freedman:2004rt}.}
What are the subdominant neutrino transitions?
Is the neutrino hierarchy normal or inverted, and what is the absolute scale of neutrino masses?
Do neutrino masses probe large extra dimensions?
How is neutrino mass a sign of physics beyond the standard model? What final conclusions can we draw from the LSND and MiniBooNE observations?
Can we find evidence for (or against) a sterile neutrino? Can we find evidence for lepton-number violation that demonstrates that neutrinos are Majorana particles? If so, do heavy right-handed ``neutrinos'' provide information about energy scales far above the electroweak scale?
Can we establish a detailed connection between neutrino mass and lepton-flavor violation?
How could light sterile neutrinos arise? Does the ``atmospheric'' mixing angle
$\theta_{23}$ correspond to maximal mixing? Is $\nu_3$ richer in $\nu_\mu$ or $\nu_\tau$? How small is $\theta_{13}$? Can we detect \textsf{CP} violation in neutrino mixing?
Does leptogenesis explain the excess of matter over antimatter in the universe?
How do neutrinos shape the universe? What constraints can we place on neutrino lifetime and on electric or magnetic dipole moments of the neutrino?

And finally, what will be the best and fastest ways to obtain information we so urgently desire about neutrinos, flavor, and identity? At the Bagn\`{e}res-de-Bigorre conference in 1953, famous as the ``last'' cosmic-ray conference before the coming of the Cosmotron at Brookhaven National Laboratory, Cecil Powell gave out the alarm, ``Gentlemen, we have been invaded! The accelerators are here.'' For his part, Louis Leprince-Ringuet offered \guillemotleft Mais nous devons aller vite, nous devons courir sans ralentir notre cadence : nous sommes poursuivis \ldots\  nous sommes poursuivis par les machines !\guillemotright\ Our situation, for the moment, is different: more and more, neutrino beams generated by particle accelerators are extending and refining knowledge gained by exploiting natural sources and reactors. As we heard during this workshop, the natural sources remain advantageous in some situations, and will benefit from larger and more capable detectors. But highly intense beams of well-defined flavor content will transform what we can do to investigate neutrino properties~\cite{Olga} and interactions~\cite{Nakaya,Petti,Yasuda,Ulrich}, and the accelerator complexes that generate them give us new possibilities to study charged-lepton physics~\cite{Sacha,Lee} and flavor physics in general. 
Moving from the current generation of neutrino detectors to a new generation, whether magnetized iron or emulsion, water Cherenkov, totally active scintillator, or liquid argon, will require both inventiveness and discipline.

The end of the workshop is a good time to examine where we want to go and how we can arrive there~\cite{scoping}. Practical considerations are sure to intervene and modulate some of the ideas we have examined. For underground experiments, cavern size, structural integrity, and excavation time all matter in the real world. Detector cost, stability, and fabrication time are no less important, and the difficulty and expense of very long baselines requiring steep dip angles is of real concern. It is not too soon to begin asking what compromises might bring us to the essential physics results in the shortest time. With respect to beta beams or a neutrino factory, we need to assess when the needed technology demonstrations be in hand, and when we will need to specify the experimental desiderata. The answers influence when it might be prudent to propose construction and how many years might be required to operations. It would be wonderful to move from planning to doing!
 
\acknowledgments
On behalf of all the participants, I want to thank our hosts from Valencia, Barcelona, and Madrid for choosing a wonderful venue, crafting a rich and stimulating scientific program, and making our time together memorably pleasant and rewarding. Anselmo Cervera deserves special mention for his energetic dedication to the cause. The Working Group leaders did much to encourage coherence and cross-pollination, speakers prepared excellent talks, and the participants contributed to vigorous discussions.  I ask forgiveness for having referred only to plenary talks.

It is my pleasure to thank Hans K\"{u}hn and Uli Nierste for splendid hospitality in Karlsruhe. I am grateful to the Alexander-von-Humboldt Stiftung for generous support. Thanks to Olga Mena for help in constructing the neutrino ternary plot in Figure~\ref{fig:flavormixing}. Next year in Batavia and Chicago!

\end{document}